# Effect of doping on the phase stability and Superconductivity in $LaH_{10}$

Zepeng Wu[1], Yang Sun[1*], Artur P. Durajski[2], Feng Zheng[1], Vladimir Antropov[3,4], Kai-MingHo[4], Shunqing Wu[1*]

[1]*Department of Physics, Xiamen University, Xiamen 361005, China*

[2]*Instiute of Physics, Częstochowa University of Technology, Ave. Armii Krajowei 19, 42-200 Częstochowa, Poland*

[3]*Ames National Laboratory, Ames, Iowa 50011, USA*

[4]*Department of Physics, Iowa State University, Ames, Iowa 50011, USA*

**Abstract**

We present a computational investigation into the effects of chemical doping with 15 different elements on phase stability and superconductivity in the $LaH_{10}$ structure. Most doping elements were found to induce softening of phonon modes, enhancing electron-phonon coupling and improving critical superconducting temperature while weakening dynamical stability. Unlike these dopants, Ce was found to extend the range of dynamical stability for $LaH_{10}$ by eliminating the van Hove singularity near the Fermi level. The doped compound, $La_{0.75}Ce_{0.25}H_{10}$, maintains high-temperature superconductivity. We also demonstrate that different Ce doping configurations in the $LaH_{10}$ structure have a minimal effect on energetic stability and electron-phonon coupling strength. Our findings suggest that Ce is a promising dopant to stabilize $LaH_{10}$ at lower pressures while preserving its high-temperature superconductivity.

[1] Email: yangsun@xmu.edu.cn (Y.S.) and wsq@xmu.edu.cn (S.W.)

## I. INTRODUCTION

In recent years, it has been experimentally observed that H-rich compounds can exhibit high-temperature superconductivity (HTS) under high pressure, such as $H_3S$ ($T_c$= 203K at 155GPa [1]), $LaH_{10}$ (~250K at 170GPa [2]; ~260K at 180-200GPa [3]), $CaH_6$ (215K at 172GPa [4,5]), $CeH_{10}$ (115K at 95GPa [6]), $CeH_9$ (~100K at 130GPa [6]), $(LaCe)H_9$ (148-178K at 97-172GPa [7,8]), $YH_9$ (243K at 201GPa [9-11]), $YH_6$ (~220K at 183GPa [9]), $(LaY)H_{10}$ (253K at 183GPa [12]) and $LaBeH_8$ (110K at 80GPa [13]). These discoveries have set a milestone in approaching the room-temperature superconductivity [14-19]. At the same time, the pressure required to stabilize these compounds is still too extreme for practical applications.

The search of binary hydrides [20,21] has shown diverse structures and chemistry in these compounds, which provide a broad platform to optimize the energetic stability and superconductivities. Compared with the binary phases, the ternary phases have a much broader configurational space, thereby offering more possibility for HTS at lower pressures [22]. It has been proposed that replacing H with small-radius elements (such as Be, B, C, N, and Si) can lower the required high pressures in the hydrides [23]. For instance, $KB_2H_8$ ( 134K-146K at 12GPa [24]), $BaSiH_8$ (71K at 3GPa [25]), $LaBH_8$ (126K at 50GPa [26]), $KPb(BC)_6$ (88K at ambient pressure [27]), $Al_2(BN)_6$ (72K at ambient pressure [28]), etc. While these dopants extend the pressure range of the stability, their superconducting temperature is simultaneously reduced.

Since the superconductivity in hydrides is mainly due to H, doping on the metal site is likely to maintain its superconductivity. Recently, high-throughput screening in the $MgB_2$-like systems shows that the doping on the metal site can effectively improve the stability and maintain the superconductivity [29]. Metals from the same family share similar characteristics, allowing them to be combined into disordered solid mixtures. This property allows us to use binary compounds as foundational blueprints for crafting ternary alloy super hydrides from the original crystal structure [30-33]. $LaH_{10}$, with the highest $T_c$ among experimentally synthesized superconductors, is a potential parent structure for doping to manipulate its HTS and pressure-dependent stability.

In this paper, based on first-principles calculations, we investigate the effects of chemical doping on phase stability and superconductivity in the $LaH_{10}$ structure. A total of 15 elements are selected as dopants: K, Rb, Cs, Ca, Sr, Ba, Sc, Y, Ti, Zr, Hf, In, Tl, Ce, and Lu. The first thirteen elements are more likely to donate electrons to H atoms to enhance the stability of the H cage framework, and the strong correlation effect caused by $d$ electrons is not significant [21]. Ce and Lu have also been theoretically predicted to have good superconducting potential [34,35]. We will use the $La_{0.75}M_{0.25}H_{10}$ model to examine their dynamical stability and superconductivity under high pressure.

## II. COMPUTATION METHODS

### 2.1 Stability calculations

The $La_{0.75}M_{0.25}H_{10}$ structure was constructed by replacing one La atom with M metal (M=K, Rb, Cs, Ca, Sr, Ba, Sc, Y, Ti, Zr, Hf, Ce, Lu, In, Tl) in the conventional cell (four formula units (f.u.)) shown in Fig. 1(a). This results in a symmetry reduction to Pm-3m. Structure relaxations and electronic properties were carried out using the Perdew-Burke-Ernzerhof (PBE) [36] functional in the framework of the projector augmented wave (PAW) method [37] as implemented in the VASP code [38]. The configurations of valence electrons used in the PAW method are shown for these elements in Table. S1. A plane-wave basis set with an energy cutoff of 500 eV and uniform Γ-centered k-point grids with a density of $2\pi \times 0.025 Å^{-1}$ were employed in the self-consistent calculations and structure relaxations. The structures were optimized until the maximum energy and force were less than $10^{-8}$ eV and 1 meV/ Å, respectively.

To investigate the dynamical stability, we used the finite displacement method by constructing a supercell with ~352 atoms and uniform Γ-centered k-point grids with a density of $2\pi \times 0.025 Å^{-1}$. The second-order force constant extraction and the harmonic phonon dispersion relationship calculation were performed with Phonopy code [39]. We employed quasi-harmonic approximation (QHA) to explore the finite temperature thermodynamics.

### 2.2 Electron-phonon coupling calculations

Harmonic phonon dispersion and electron-phonon coupling (EPC) were calculated within the density functional perturbation theory (DFPT) [40], as implemented in the QUANTUM ESPRESSO package [41,42]. Ultrasoft pseudopotentials [43] with PBE functional were used with a kinetic energy cutoff of 80 Ry and a charge density cutoff of 800 Ry. The valence electron configurations used in USPP were the same as in PAW potential, so the calculations performed with QE and VASP were consistent. Self-consistent electron density and EPC were calculated by employing 8×8×8 $k$-point meshes and 4×4×4 $q$-point meshes. A dense 16×16×16 $k$-point mesh was used for evaluating electron-phonon interaction matrix.

The main input element to the Eliashberg equations is the Eliashberg spectral equation $\alpha^2 F(\omega)$ defined as[44,45]

$$\alpha^2 F(\omega) = \frac{1}{2\pi N(E_F)} \sum_{qv} \frac{\gamma_{qv}}{\hbar\omega_{qv}} \delta(\omega - \omega_{qv}) \quad (1)$$

where $N(E_F)$ is the states at the Fermi level $E_F$, $\omega_{qv}$ representative the phonon frequency of the mode $v$ with wave vector $q$. The phonon linewidth $\gamma_{qv}$, which is the imaginary part of the phonon self-

energy, is defined as

$$\gamma_{qv} = \frac{2\pi\omega_{qv}}{\Omega_{B.Z}} \sum_{i,j} \int d^3k\, |g_{k,qv}^{i,j}|^2 \delta(\varepsilon_{i,q} - E_F)\delta(\varepsilon_{j,k+q} - E_F) \tag{2}$$

$g_{k,qv}^{i,j}$ is the EPC matrix element, and $\Omega_{B.Z}$ is the volume of the Brillouin zone (B.Z.).The EPC constant is calculated by

$$\lambda = \sum_{qv} \frac{\gamma_{qv}}{\pi\hbar N(E_F)\omega_{qv}^2} = 2\int_0^{\infty} \frac{\alpha^2 F(\omega)}{\omega} d\omega \tag{3}$$

We chose the gaussian smearing width of 0.02-0.03 Ry based on the convergence test in Supplementary Note 1. $T_c$ was first estimated using McMillan-Allen-Dynes (MAD) formula [44,45] with Coulomb pseudopotential $\mu^*$= 0.13 [46,47].

$$T_c = \frac{f_1 f_2 \omega_{log}}{1.2} exp\left(-\frac{1.04(1+\lambda)}{\lambda - \mu^*(1+0.62\lambda)}\right) \tag{4}$$

where $f_1$ $and$ $f_2$ are two separate correction factors [44], which are functions of $\lambda$, $\omega_{log}$, $\omega_2$, and $\mu^*$. The logarithmic average frequency $\omega_{log}$ is computed as:

$$\omega_{log} = exp\left(\frac{2}{\lambda}\int_0^{\infty} \frac{\alpha^2 F(\omega)}{\omega} \ln\omega\, d\omega\right) \tag{5}$$

**2.3 Migdal-Eliashberg approach**

The thermodynamic properties of superconducting ternary $La_{0.75}M_{0.25}H_{10}$ hydrides were also estimated using the Migdal-Eliashberg (ME) approach due to the strong electron-phonon coupling constants observed in these systems. The isotropic Eliashberg equations defined on the imaginary-frequency axis, which incorporate the superconducting order parameter function $\varphi_n = \varphi(i\omega_n)$ and the electron mass renormalization function $Z_n = Z(i\omega_n)$ take the following form [48,49]:

$$\varphi_n = \frac{\pi}{\beta} \sum_{m=-M_f}^{M_f} \frac{\lambda_{n,m} - \mu^*\theta(\omega_c - |\omega_m|)}{\sqrt{\omega_m^2 Z_m^2 + \varphi_m^2}} \varphi_m, \tag{6}$$

$$Z_n = 1 + \frac{1}{\omega_n}\frac{\pi}{\beta} \sum_{m=-M_f}^{M_f} \frac{\lambda_{n,m}}{\sqrt{\omega_m^2 Z_m^2 + \varphi_m^2}} \omega_m Z_m, \tag{7}$$

where $\beta = 1/k_B T$, and the electron-phonon interaction pairing kernel is given by,

$$\lambda_{n,m} = 2\int_0^{\infty} \frac{\omega}{(\omega_n - \omega_m)^2 + \omega^2} \alpha^2 F(\omega) d\omega. \tag{8}$$

Hence, the superconducting order parameter was defined by the ratio $\Delta_n = \varphi_n / Z_n$ and the superconducting transition temperature $T_c$ was estimated from the following relation $\Delta_{n=1}(\mu^*, T =$ $T_c$) = 0. We used the same Coulomb pseudopotential as the one used in MAD calculations, i.e., $\mu^*$=0.13. The Eliashberg equations were solved iteratively in a self-consistent way with a maximal error of $10^{-10}$ between two successive iterations. The convergence was controlled by the sufficiently high number of Matsubara frequencies: $\omega_n = (\pi/\beta)(2n-1)$, where $n = 0, \pm 1, \pm 2, \ldots, \pm M_f$ and $M_f$ = 1100 [50-52].

## III. RESULTS AND DISCUSSION

### 3.1 Phase stability

We first evaluate the dynamical stability of ternary $La_{0.75}M_{0.25}H_{10}$ structures. Harmonic phonon dispersions were calculated for all 16 phases at 400GPa, 250GPa, and 200GPa (see Supplementary Fig. S5). A phase without any imaginary modes in the phonon spectrum is marked as dynamically stable in Fig. 1(b). At 400 GPa, the structure is stable with seven substitutions, i.e., Sr, Ba, Y, Zr, Hf, Ce, and Lu. Y and Ce substitutions can maintain stability when the pressure is reduced to 250 GPa. At 200 GPa, only $La_{0.75}Ce_{0.25}H_{10}$ remains stable at harmonic level. $LaH_{10}$ becomes harmonic dynamically unstable below 230 GPa (see Fig. S6). Therefore, Ce substitution can improve the stability of $LaH_{10}$ and lower the pressure range of the stability. Our calculations were based on the harmonic approximation, while the anharmonic effect and the quantum nuclear effect (QNE) were ignored. The anharmonic oscillations of the hydrogen sublattice can contribute to the $T_c$ and thermodynamic stability of hydrides [53-56]. The calculations with QNE and anharmonic correction indicate the $LaH_{10}$ can be stabilized as low as ~ 130 GPa [53,57], similar to the experimental observation at ~ 140 GPa [58]. Therefore, the pressure stability range of present $La_{0.75}Ce_{0.25}H_{10}$ is expected to expand further by including anharmonic and QNE effects.

Given the harmonic dynamical stability, we evaluate the thermodynamic stability of $La_{0.75}Ce_{0.25}H_{10}$. We calculated its enthalpy on the ternary phase diagram at 200 GPa, as shown in Fig. S2(a). The results show that the energy of the $La_{0.75}Ce_{0.25}H_{10}$ structure is only 1 meV/atom higher than that of the convex hull. In addition, we also considered finite temperature thermodynamics (see Supplementary Note 2) and found the of $La_{0.75}Ce_{0.25}H_{10}$ (Pm-3m) has promising thermodynamic stability up to 300 K.

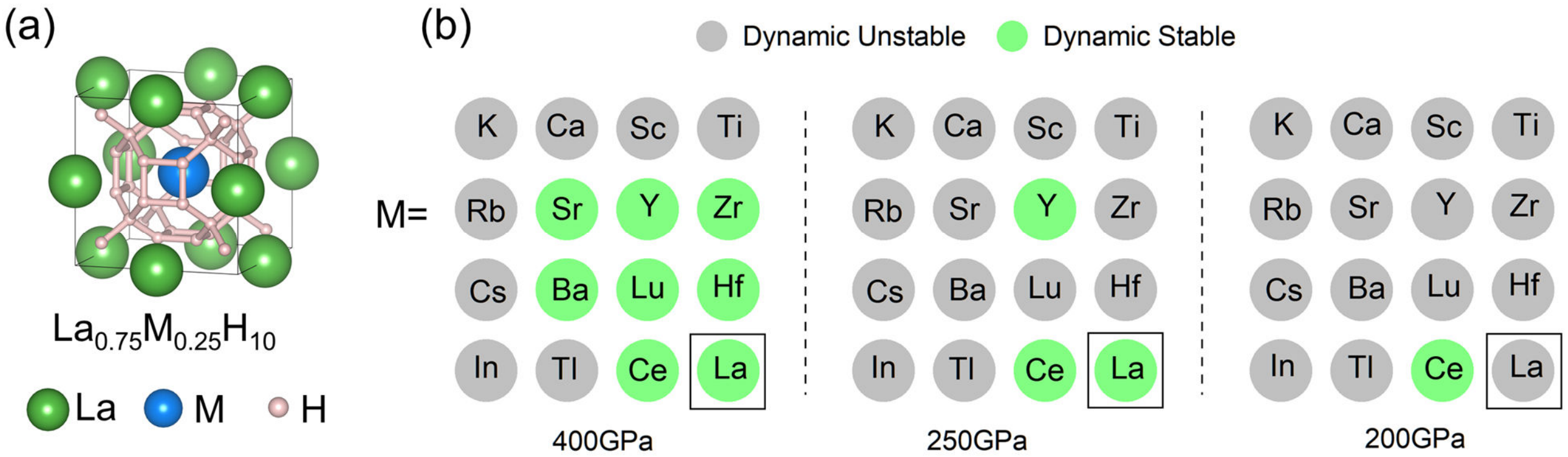

Fig. 1. (a) Structure of $La_{0.75}M_{0.25}H_{10}$, M=K, Rb, Cs, Ca, Sr, Ba, Sc, Y, La, Ti, Zr, Hf, In, Tl, Ce, Lu. (b) Dynamic stability of all doped phases at 400 GPa, 250 GPa, and 200 GPa.

### 3.2 Electron-phonon coupling and superconductivity

We calculate the EPC constant λ using the DFPT method and Eliashberg theory for the dynamically stable structures at 400, 250, and 200GPa. We first compute the superconducting transition temperature ($T_c$) by the MAD formula, presented in Table 1. Due to the large λ (>2) in these compounds, we also employ Eliashberg formalism to investigate the impact of EPC on the $T_c$ and superconducting energy gap. The temperature-dependent behavior of the superconducting energy gap $\Delta(T)$ is computed by solving the ME equations in the mixed representation (defined simultaneously on the imaginary and real axis) [59,49]. The results are presented in Fig. 2, which illustrates the calculated Δ(T) curves for dynamically stable structures of $La_{0.75}M_{0.25}H_{10}$ at 400 GPa.

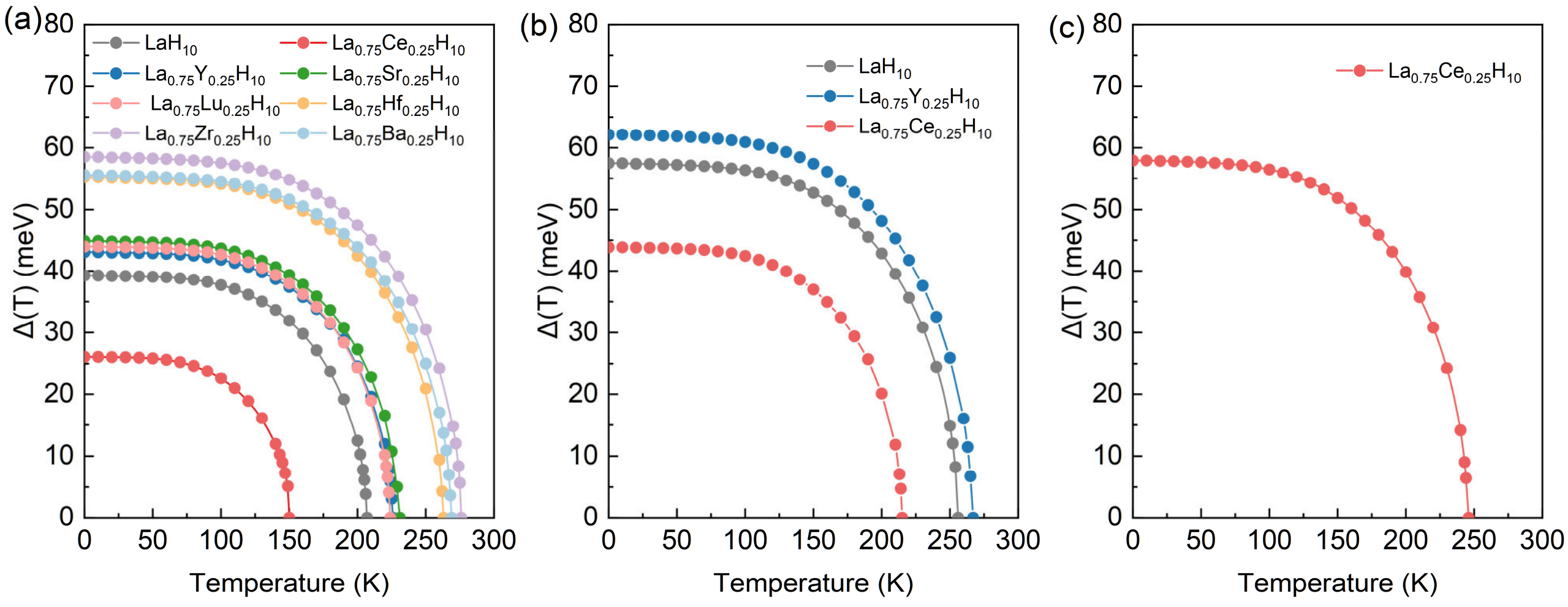

Fig. 2. Superconducting energy gap as a function of temperature for $La_{0.75}M_{0.25}H_{10}$ at (a) 400GPa, (b) 250GPa, and (c) 200GPa.

Based on Δ(T) results, we estimate $T_c$ and compare it with MAD results in Table 1. $T_c$ is found to be high for all investigated cases and reaches the highest value of 276 K for $La_{0.75}Zr_{0.25}H_{10}$ at 400 GPa and 267 K for $La_{0.75}Y_{0.25}H_{10}$ at 250 GPa. The $T_c$ values of $La_{0.75}M_{0.25}H_{10}$ predicted via the MAD formula are consistently lower (underestimated) than those obtained from the ME formalism, particularly for the one with large $\lambda$. This justifies the usage of the ME formalism: we assumed an underestimation of $T_c$ in the MAD method using the strong coupling ME method. The obtained results entirely confirm the assumption. The calculation of $LaH_{10}$ shows that $\lambda$ is 2.53 and $T_c$ is 256 K at 250 GPa by ME equation. As a reference, the experimental $T_c$ of $LaH_{10}$ was observed at ~250K under 170-200 GPa. Therefore, our calculation of $T_c$ is consistent with the experimental data. Below, we use $T_c$ from ME formalism for further analysis.

TABLE I. Superconducting critical temperature ($T_c$) of dynamically stable $La_{0.75}M_{0.25}H_{10}$ at 200, 250, and 400GPa estimated using Migdal-Eliashberg approach $T_{c_ME}$.and MAD formula $T_{c_MAD}$

| **P(GPa)** | **Structure** | $\boldsymbol{\lambda}$ | $\boldsymbol{T_{c_ME}}$ **(K)** | $\boldsymbol{T_{c_MAD}}$ **(K)** |
|---|---|---|---|---|
| **200** | $La_{0.75}Ce_{0.25}H_{10}$ | 3.08 | 246 | 209 |
| **250** | $LaH_{10}$ | 2.53 | 256 | 220 |
| | $La_{0.75}Ce_{0.25}H_{10}$ | 1.83 | 215 | 186 |
| | $La_{0.75}Y_{0.25}H_{10}$ | 3.16 | 267 | 208 |
| **400** | $LaH_{10}$ | 1.41 | 207 | 174 |
| | $La_{0.75}Ce_{0.25}H_{10}$ | 1.07 | 150 | 125 |
| | $La_{0.75}Y_{0.25}H_{10}$ | 1.55 | 226 | 188 |
| | $La_{0.75}Sr_{0.25}H_{10}$ | 1.69 | 231 | 186 |
| | $La_{0.75}Lu_{0.25}H_{10}$ | 1.73 | 224 | 181 |
| | $La_{0.75}Hf_{0.25}H_{10}$ | 2.32 | 263 | 190 |
| | $La_{0.75}Zr_{0.25}H_{10}$ | 2.34 | 276 | 210 |
| | $La_{0.75}Ba_{0.25}H_{10}$ | 2.34 | 269 | 178 |

In Fig. 3(a), we found that substitution with Y, Sr, Lu, Hf, Zr, and Ba all enhance the EPC constant and $T_c$ at 400 GPa, while the substitution with Ce weakens them. Similarly, at 250 GPa, $\lambda$ and $T_c$ increase with Y substitution while decreasing with Ce substitution. At 200 GPa, the only stable phase $La_{0.75}Ce_{0.25}H_{10}$ remains a potential high-$T_c$ superconductor with $T_c$ of 246 K and $\lambda$ of 3.08.

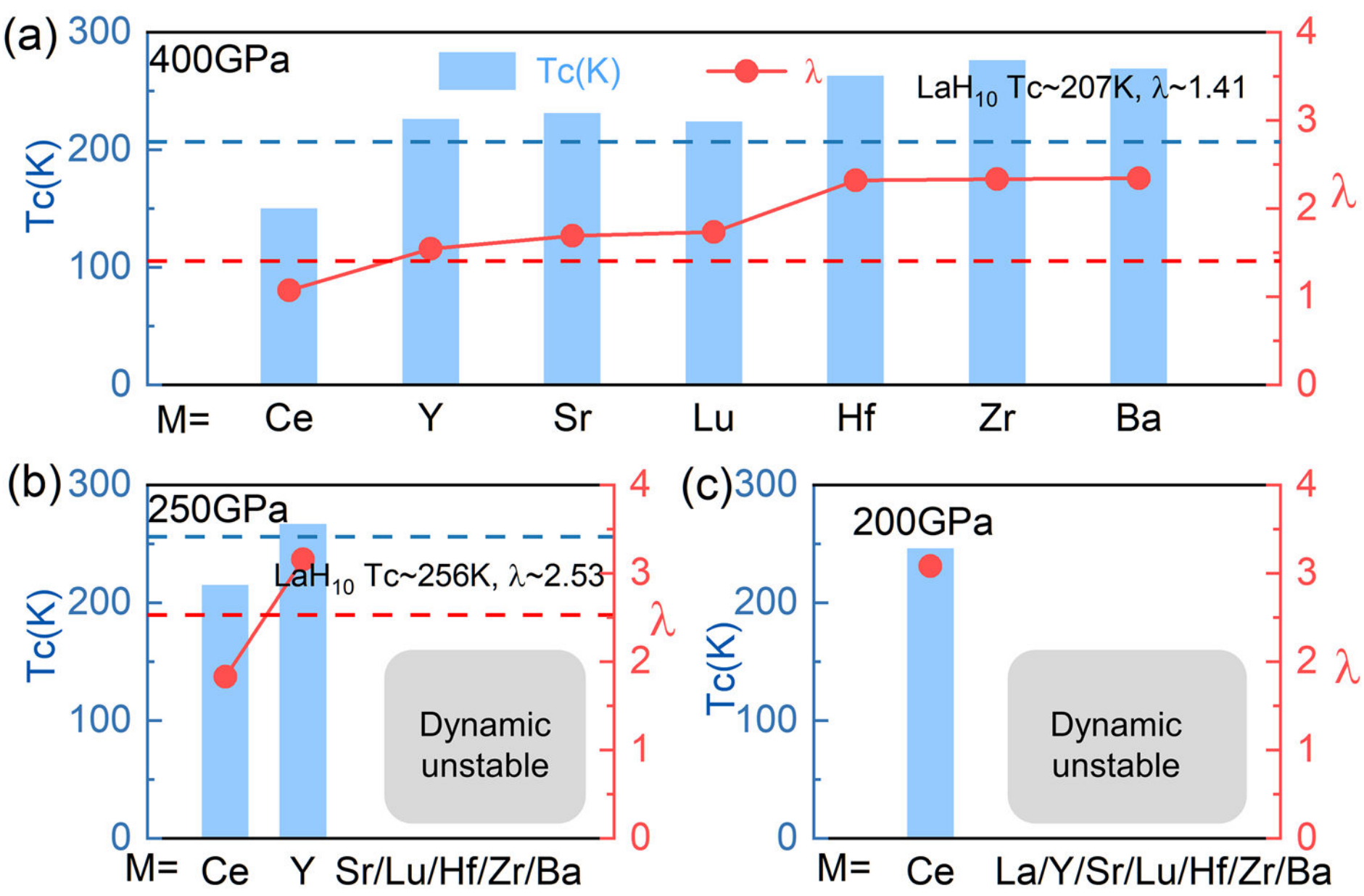

Fig. 3. Superconducting transition temperature ($T_c$) with and electron-phonon coupling constant λ of stable $La_{0.75}M_{0.25}H_{10}$ structures at (a) 400GPa, (b) 250GPa and (c) 200GPa

To understand the origin of the increased λ and $T_c$ by doping, we use $La_{0.75}Hf_{0.25}H_{10}$ as an example and compare its phonon spectra to the $LaH_{10}$ in Fig. 4. We find the substitution of La with Hf induces significant softening of high-frequency phonon modes. As shown in Fig. 4(a), with the Hf substitution, a few phonon modes appear in the low-frequency range of 360-900 cm$^{-1}$, while no phonon modes exist in the same area for $LaH_{10}$. The H atoms dominate these phonon modes (see the projected phonon DOS in Fig. S7). Comparing the Eliashberg spectral function between $LaH_{10}$ and $La_{0.75}Hf_{0.25}H_{10}$ in Fig. 4 (b) and (c), one can see the phonon softening at the range of 360-900 cm$^{-1}$ significantly promotes the EPC in this region. Similar enhancement of phonon linewidth in 360-900 cm$^{-1}$ can be found by comparing Fig. 4 (d) and (e). If we integrate Eq. (3) to $\omega = 900\text{cm}^{-1}$, we find the contribution to $\lambda$ from frequencies less than $900\text{cm}^{-1}$ is 0.18 and 1.01 for $LaH_{10}$ and $La_{0.75}Hf_{0.25}H_{10}$, respectively. Therefore, the phonon softening in $La_{0.75}Hf_{0.25}H_{10}$ significantly enhances the EPC. This mechanism is also seen in other superconducting systems [60-63]. The analysis of $La_{0.75}Hf_{0.25}H_{10}$ illustrates that substituting La with Hf changes the bonding with H atoms and softens vibrational modes. Such phonon softening enhances the EPC and increases the λ and $T_c$, simultaneously. We also analyzed the EPC for other dopants and found similar effects, as shown in Fig. S8 and Table S2, i.e., the substitution of La leads to phonon softening, which contributes to strong EPC in the middle- and low-frequency regions.

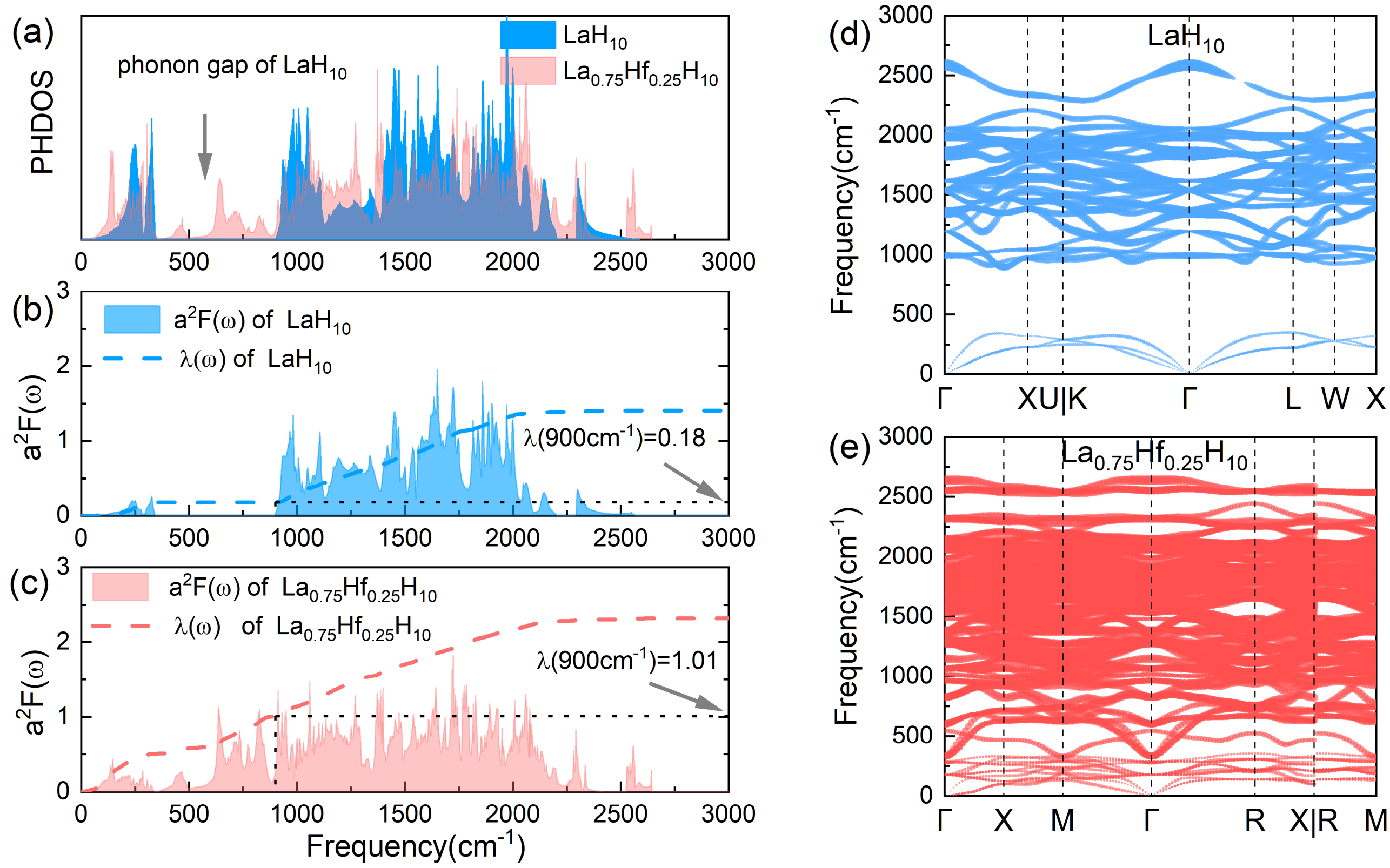

Fig. 4. (a) phonon dos of $LaH_{10}$ and $La_{0.75}Hf_{0.25}H_{10}$ at 400GPa. (b) and (c) Eliashberg spectrum function $\alpha^2F(\omega)$, and electron-phonon coupling integral $\lambda(\omega)$ of $LaH_{10}$ and $La_{0.75}Hf_{0.25}H_{10}$ at 400GPa. (d) and (e) Phonon spectrum of $LaH_{10}$ and $La_{0.75}Hf_{0.25}H_{10}$ at 400GPa. The solid circles show the EPC with the area proportional to the respective phonon linewidth.

### 3.3 The effects of Ce

Ce is the only substitution that increases the pressure range of $LaH_{10}$ stability while maintaining the high-temperature superconductivity with a slight weakening of the EPC in the harmonic approximation. To understand the effect of Ce substitution on dynamic stability, we compare the phonon spectrum between $LaH_{10}$ and $La_{0.75}Ce_{0.25}H_{10}$ at 200 GPa in Fig. 5(a) and (b). In $LaH_{10}$, the imaginary frequency modes on the Γ-X, Γ-M, and Γ-R paths are dominated by the vibrations of hydrogen atoms. When Ce is introduced, these modes become stiffer, and the imaginary frequency disappears. In Fig. 5(c) and (d), we compare the electronic band structure and density of states for $LaH_{10}$ and $La_{0.75}Ce_{0.25}H_{10}$, respectively. $LaH_{10}$ shows a flat band near the Fermi level with eightfold degeneracy at the **R** point. This caused a Van Hove singularity (VHS) in the density of states. Ce substitution opens the gap at **R** and splits the degenerated bands. This removes the VHS and reduces the states at the Fermi level. Correspondingly, the imaginary modes at **R** disappear.

Moreover, additional bands contributed mainly by Ce and H cross the Fermi level at Γ-M and Γ-R paths in $La_{0.75}Ce_{0.25}H_{10}$. The bonding likely contributes to the hardening of phonon modes. Based on the electronic density of states, these Ce bands near the Fermi level are mostly from 4*f* orbitals. Therefore, this indicates that the 4*f* electron in Ce contributes significantly to the dynamic stability of $La_{0.75}Ce_{0.25}H_{10}$. To further validate this mechanism, we computed the phonon spectrum of $La_{0.75}Ce_{0.25}H_{10}$ with the ultrasoft pseudopotential where Ce's 4*f* electrons are regarded as core electrons. This ultrasoft pseudopotential leads to charge transfer and the re-appearance of imaginary modes caused by the Ce-4*f* electron as discussed in Supplementary Note 3. These results suggest the strong effect of Ce-4*f* electrons in stabilizing the $LaH_{10}$ at low pressures.

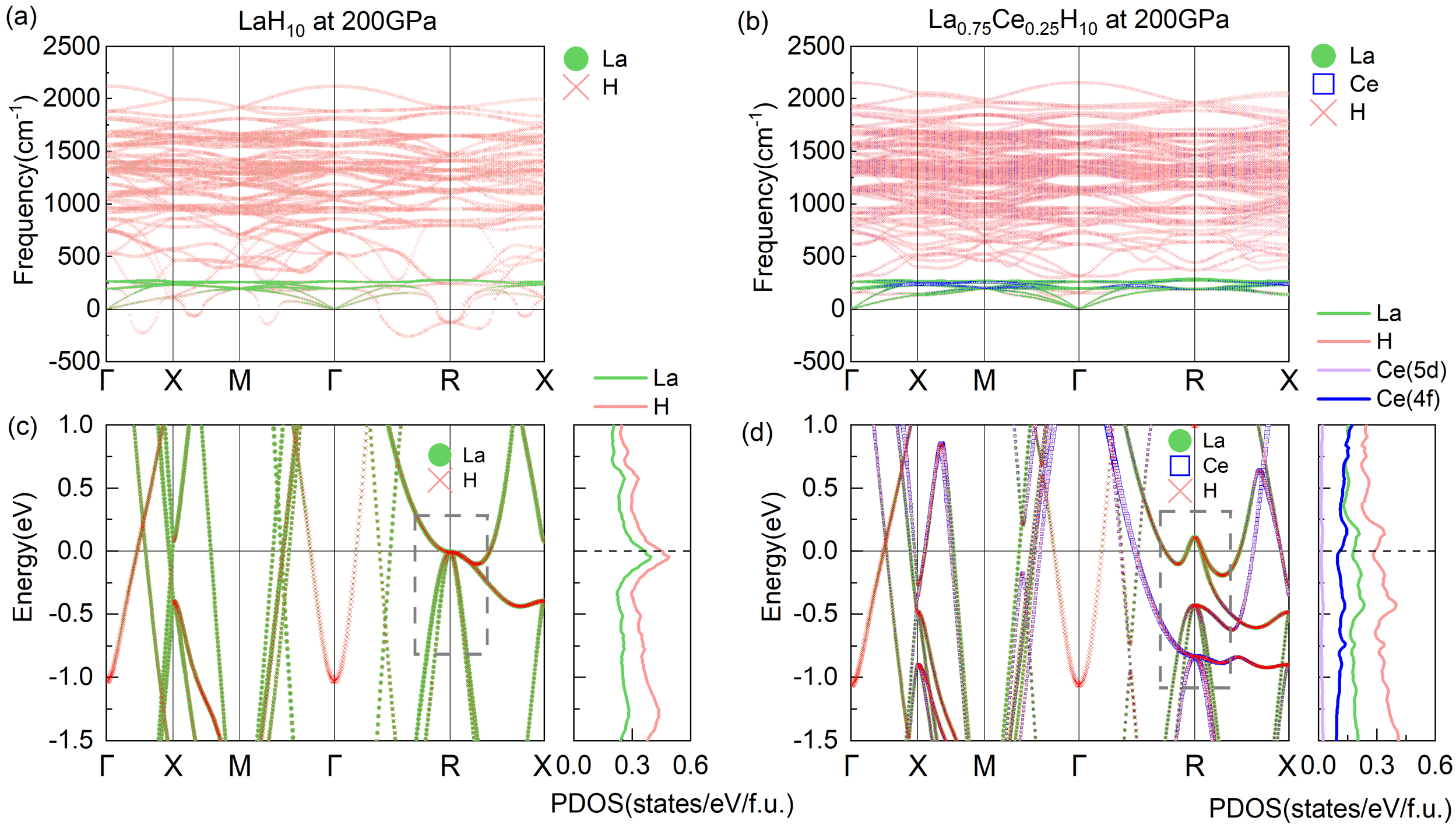

Fig. 5. (a) (b) Atom-projected phonon spectrum of $LaH_{10}$ and $La_{0.75}Ce_{0.25}H_{10}$ at 200GPa. (c) (d) fat electron band (and projected density of states, PDOS) of $LaH_{10}$ and $La_{0.75}Ce_{0.25}H_{10}$ at 200GPa.

So far, the substitutional effect of Ce was only considered with Pm-3m $La_{0.75}Ce_{0.25}H_{10}$ structure. We further examine the stability of other $La_{0.75}Ce_{0.25}H_{10}$ polymorphs at 200 GPa. As shown in Fig. 6, we construct $LaH_{10}$ supercells (88 atoms) by $2 \times 2 \times 2$, $1 \times 1 \times 8$ and $1 \times 2 \times 4$ and randomly replace La atoms with Ce atoms to generate 9 unique structures. Energy calculations show that these structures all have similar enthalpy with differences less than 8 meV/atom. Harmonic phonon calculations shown in Fig. S9 suggest five phases are dynamically stable, which is noted in Fig. 6. To explore the possible superconductivity in these structures, we employ a recently developed frozen-phonon method to compute the zone-center EPC strength for stable structures. This efficient method can identify strong EPC candidates in hydrides because the zone-center EPC strongly correlates with the full Brillouin zone EPC in these materials [64]. Using this method, we compute the zone-center EPC, $\lambda_\Gamma$, for 5 dynamically stable polymorphs. As shown in Fig. 6, different structures show similar $\lambda_\Gamma$ as the one of the Pm-3m phase. Therefore, Ce occupation in the $La_{0.75}Ce_{0.25}H_{10}$ does not affect its energetic stability and EPC. To confirm the zone-center EPC calculations, we also performed DFPT calculations of full Brillouin zone EPC for the P4/mmm phase (see details in Fig. S10). We obtained $\lambda$ of P4/mmm as 2.64, slightly smaller than the Pm-3m phase ($\lambda$=3.08). This is consistent with the zone-center EPC calculations. The $T_c$ was estimated 215K (with ME approach) at 200GPa, which is slightly smaller than the one of Pm-3m phase (246K). Since these polymorphs have similar energy, they may form a random solid solution in the experimental synthesis. Nevertheless, such a mixture should maintain the HTS because of the similar electron-phonon coupling strength in these phases.

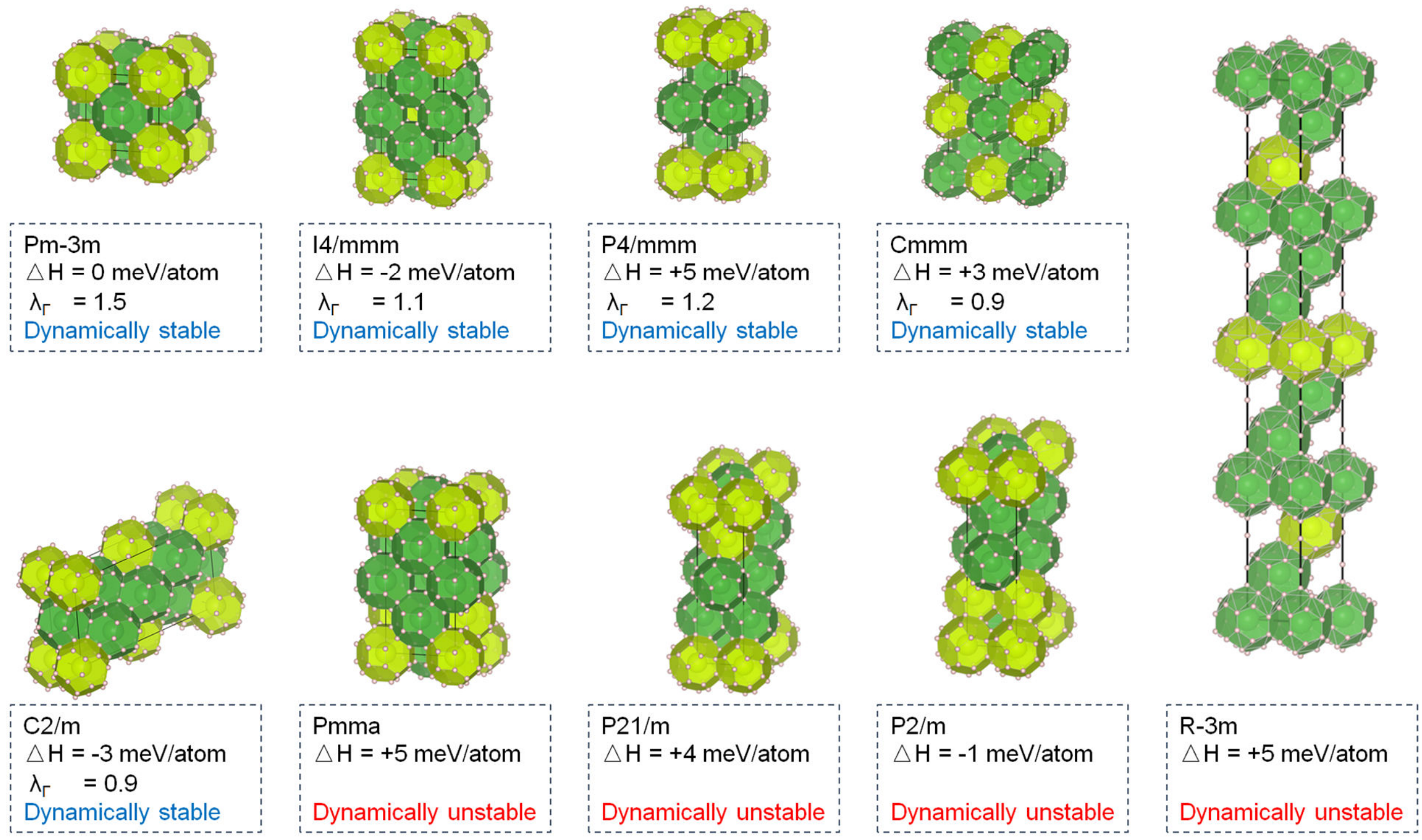

Fig. 6. The crystal structure, relative enthalpy ΔH and zone-center EPC $\lambda_\Gamma$ of 9 $La_{0.75}Ce_{0.25}H_{10}$ polymorphs at 200GPa. The green (yellow) polyhedron represents La-H (Ce-H) cages.

Additional effects such as spin-orbit coupling (SOC) and electron correlation of *f*-electron in Ce may affect the superconductivity of $La_{0.75}Ce_{0.25}H_{10}$. However, calculating the EPC and $T_c$ directly under these effects is highly complex and sophisticated. Therefore, we performed additional SOC and DFT+U calculations to understand their effect on the electronic band structure and phonon dispersion spectrum instead of direct calculations of EPC. Here, we choose the U (Ce-4*f*) value of 4 eV [65]for the PBE+U calculation. As shown in Fig. 7, both SOC and DFT+U calculations result in electronic and phonon band structures similar to the one without these effects. Therefore, we expect these effects should be weak on the EPC of $La_{0.75}Ce_{0.25}H_{10}$.

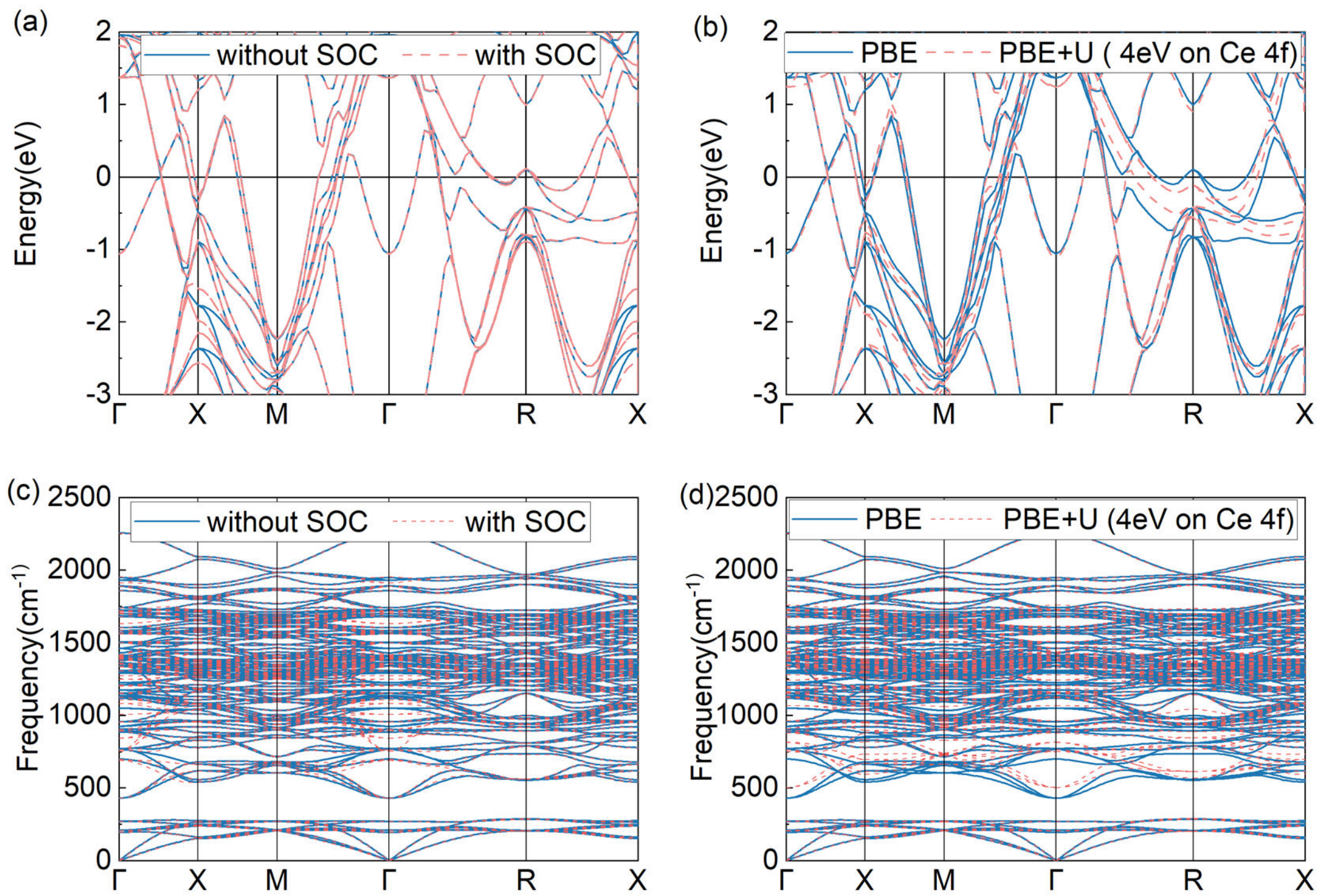

Fig. 7 (a) (b) Electron band structure of $La_{0.75}Ce_{0.25}H_{10}$ with or without SOC or U effects. (c) (d) phonon spectrum, respectively.

## IV. CONCLUSIONS

In summary, based on first-principles calculations, we have investigated the effects of chemical doping on phase stability and superconductivity in the $LaH_{10}$ structure. By analyzing the phonon spectrum, we demonstrated that most doping elements (K, Rb, Cs, Ca, Sr, Ba, Sc, Y, Ti, Zr, Hf, Lu, In, Tl) induce the softening of the high-frequency phonon modes, thereby enhancing the EPC and improving $T_c$. However, phonon softening also leads to dynamical instability, reducing the stable pressure range. Unlike these dopants, Ce doping can expand the range of dynamical stability for $LaH_{10}$ at lower pressures. The analysis of the electronic structures revealed that Ce doping eliminates the VHS and reduces states at the Fermi level, stiffening a few imaginary modes in $LaH_{10}$ at low pressures. Utilizing the Eliashberg theory, we demonstrated that $La_{0.75}Ce_{0.25}H_{10}$ maintains high-temperature superconductivity with a $T_c$ of ~ 246K at 200GPa. Upon examining different polymorphs of $La_{0.75}Ce_{0.25}H_{10}$, we show that different doping sites of Ce in the $LaH_{10}$ structure have a minor effect on the energetic stability and EPC. Our findings suggest Ce can be a promising dopant to stabilize $LaH_{10}$ at lower pressures while preserving its high-temperature superconductivity. The experimental verification of our prediction is highly desirable.

## Acknowledgments

Y.S. acknowledges support from the Fundamental Research Funds for the Central Universities (20720230014). V.P. was supported by the U.S. Department of Energy, Office of Basic Energy Sciences, Division of Materials Sciences and Engineering. Ames National Laboratory is operated for the U.S. Department of Energy by Iowa State University under Contract No. DE-AC02-07CH11358. K.M.H. acknowledges support from National Science Foundation Awards No. DMR-2132666. A.P.D. acknowledges financial support from the National Science Centre (Poland) under project No. 2022/47/B/ST3/0062. S. Fang and T. Wu from the Information and Network Center of Xiamen University are acknowledged for their help with GPU computing. Tan Kah Kee Supercomputing Center is acknowledged for its support of high-performance computing.

**Supplementary Material**

# Effect of Doping on the phase stability and Superconductivity in $LaH_{10}$

Zepeng Wu[1], Yang Sun[1*], Artur P. Durajski[2], Feng Zheng[1], Vladimir Antropov[3,4], Kai-MingHo[4], Shunqing Wu[1*]

*[1]Department of Physics, Xiamen University, Xiamen 361005, China*

*[2]Instiute of Physics, Częstochowa University of Technology, Ave. Armii Krajowei 19, 42-200 Częstochowa, Poland*

*[3]Ames National Laboratory, Ames, Iowa 50011, USA*

*[4]Department of Physics, Iowa State University, Ames, Iowa 50011, USA*

This pdf contains:

- Supplementary Note 1-3
- Supplementary Figures S1-S10
- Supplementary Tables S1-S2
- Supplementary References

[1] Email: yangsun@xmu.edu.cn (Y.S.) and wsq@xmu.edu.cn (S.W.)

**Supplementary Note 1** | Convergence test of electron-phonon coupling calculations

We test the convergence of lambda with different smearing coefficients using the method suggested in Ref. [1-3]. As shown in Fig.S1, considering the same precision for all structures ($La_{0.75}M_{0.25}H_{10}$ with lattice constant ~ 4.5Å), we believe that broadening of ~ 0.02-0.03Ry would be a suitable selection.

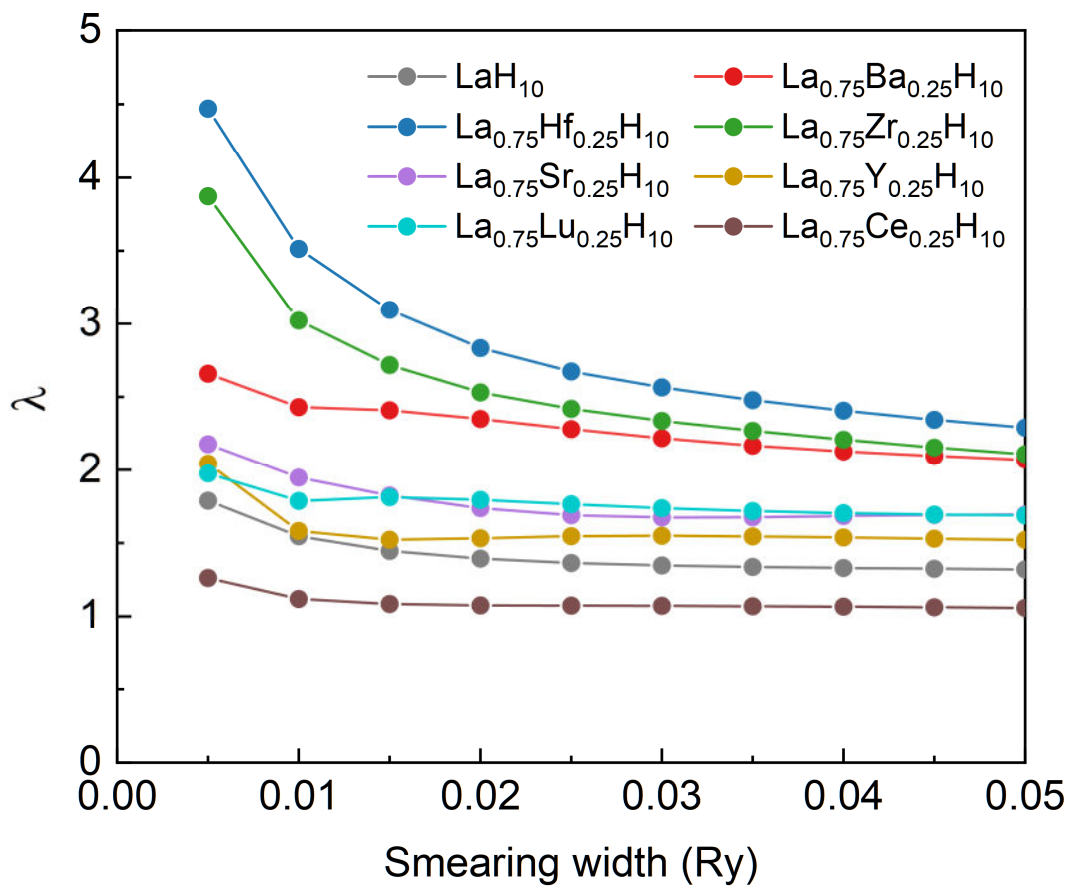

Fig. S1. λ dependence on the gaussian smearing width in $La_{0.75}M_{0.25}H_{10}$ compounds.

**Supplementary Note 2** | Finite-temperature thermodynamic stability of $La_{0.75}Ce_{0.25}H_{10}$.

We analyze the finite-temperature thermodynamic effect for La-Ce-H. The convex hull in Fig. S2 (a) indicates the competing phases with $La_{0.75}Ce_{0.25}H_{10}$ are $LaH_{10}$ and $La_{0.5}Ce_{0.5}H_{10}$. Therefore, we consider the finite-temperature thermodynamics by computing the Gibbs free energy change of the reaction $0.5LaH_{10}+0.5La_{0.5}Ce_{0.5}H_{10} \rightarrow La_{0.75}Ce_{0.25}H_{10}$. We employed quasi-harmonic approximation (QHA) to compute the Gibbs free energy for these phases. With static calculations, $La_{0.75}Ce_{0.25}H_{10}$ (Pm-3m) is above the convex hull by 1 meV/atom. From Fig. S2 (b), we find the free energy difference does not change significantly with the temperature. Therefore, the vibrational effect does not show a strong contribution on the stability of $La_{0.75}Ce_{0.25}H_{10}$.

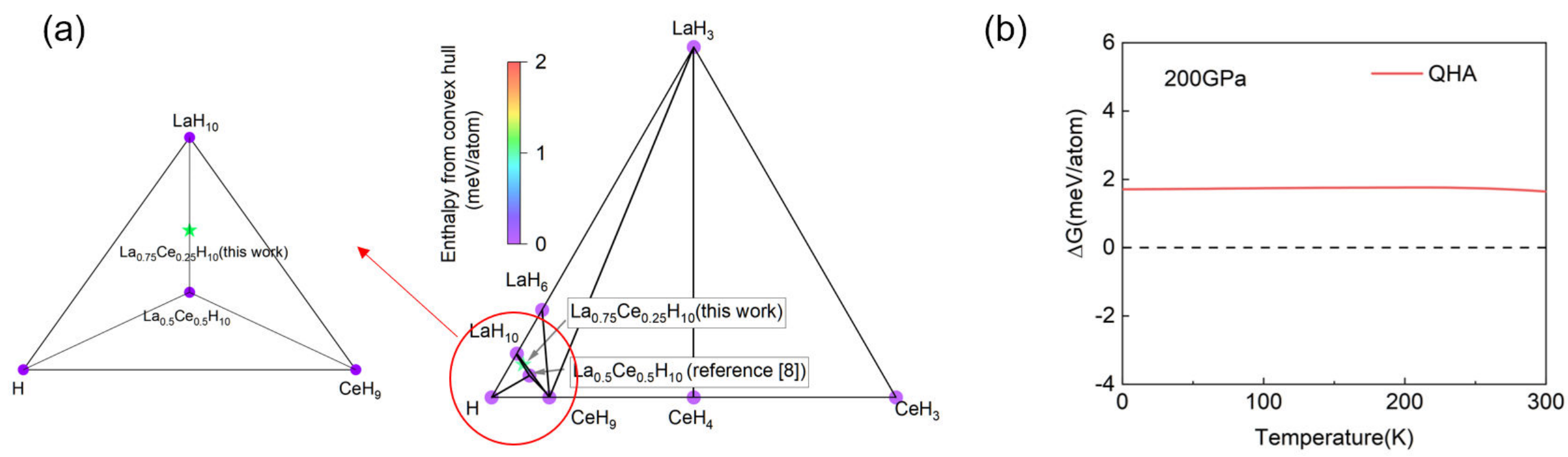

Fig. S2. (a) Convex hull without ZPE of La-Ce-H system at 200GPa. Reference structures were select from [4-8]. (b) Relative Gibbs free energy difference at 200GPa for $0.5LaH_{10}+0.5La_{0.5}Ce_{0.5}H_{10} \rightarrow La_{0.75}Ce_{0.25}H_{10}$.

**Supplementary Note 3** | Effect of *4f* electron

To study the effect of 4*f* electron on the stability of $La_{0.75}Ce_{0.25}H_{10}$, we compute phonon spectrum with and without 4*f* electrons. Figure S3 shows that inclusion of Ce's 4*f* electron can stabilize the imaginary frequency phonon modes.

To understand this effect, we analyzed the spatial difference of charge density distribution between calculations with and without *4f* electrons in Fig. S4(a). With the presences of Ce-*4f* electrons, the charge near H atoms forming the Ce cage are significantly reduced (blue region). The Bader charge analysis also shows a charge decrease of ~0.005-0.015 per H atom in the Ce cage. This can be due to the significant increase of charge near Ce atoms with Ce-4*f* electron inclusion. The change of charge distribution modifies the partial density of states. As shown in Fig. S4(b), with the Ce-4*f* electron the states of H-1*s* at the Fermi level are reduced. All these leads to the stabilization of the phonon spectrum.

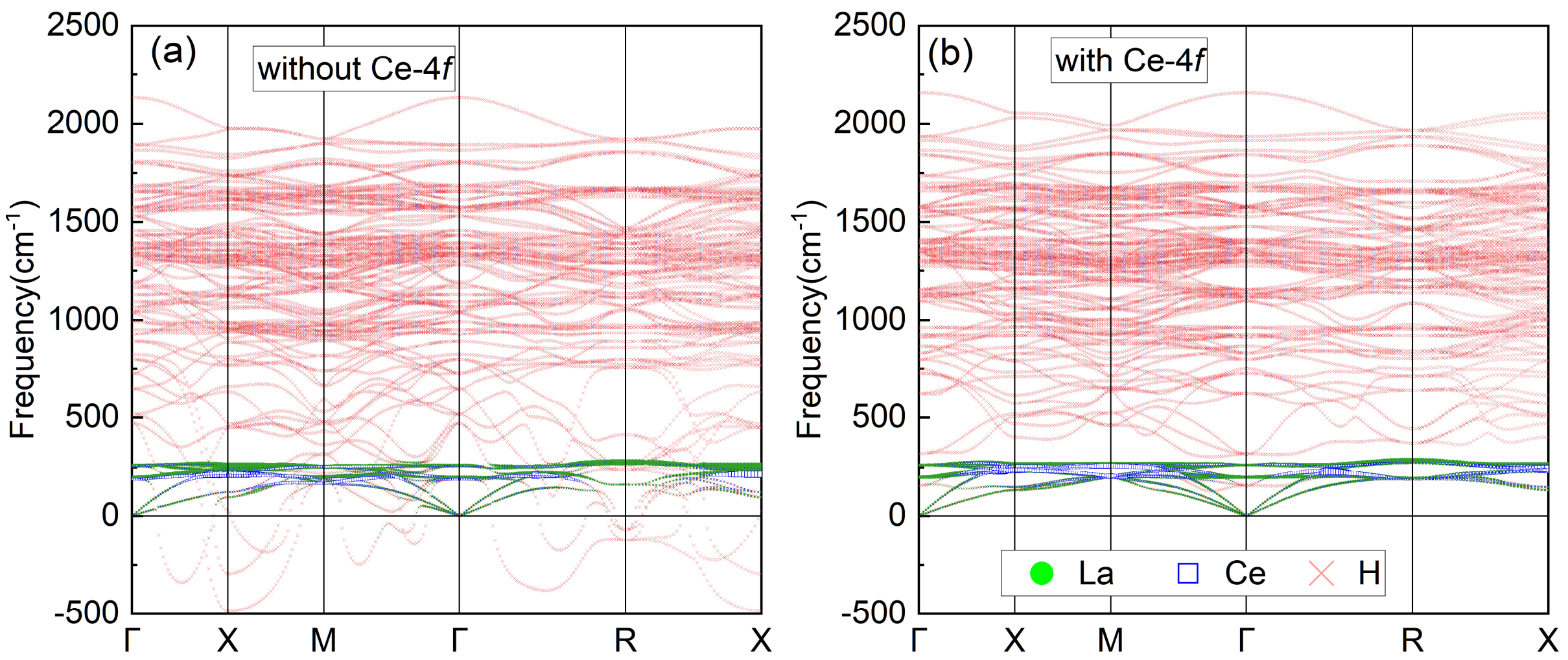

Fig. S3. Atom-projected phonon spectrum of $La_{0.75}Ce_{0.25}H_{10}$ at 200GPa without (a) and with (b) Ce-4*f* electron at 200GPa.

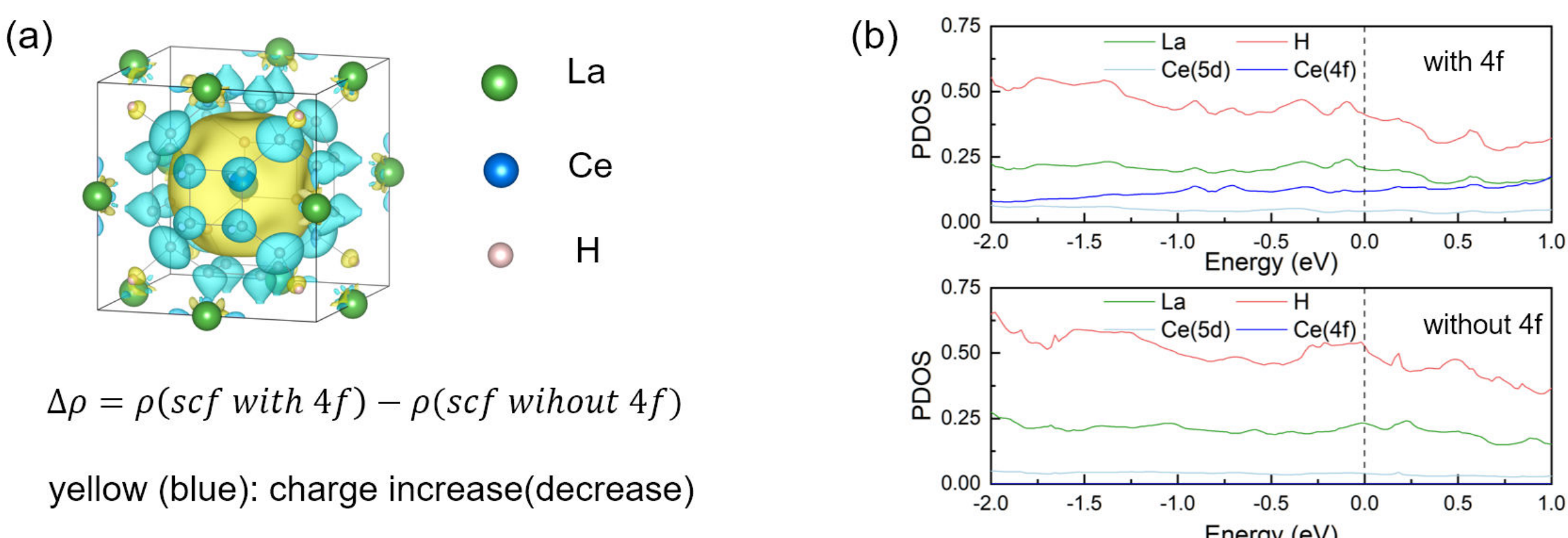

Fig. S4. (a) Charge density difference with and without the inclusion of Ce-4*f* electron. (b) Partial density of states with and without Ce-4*f* electron.

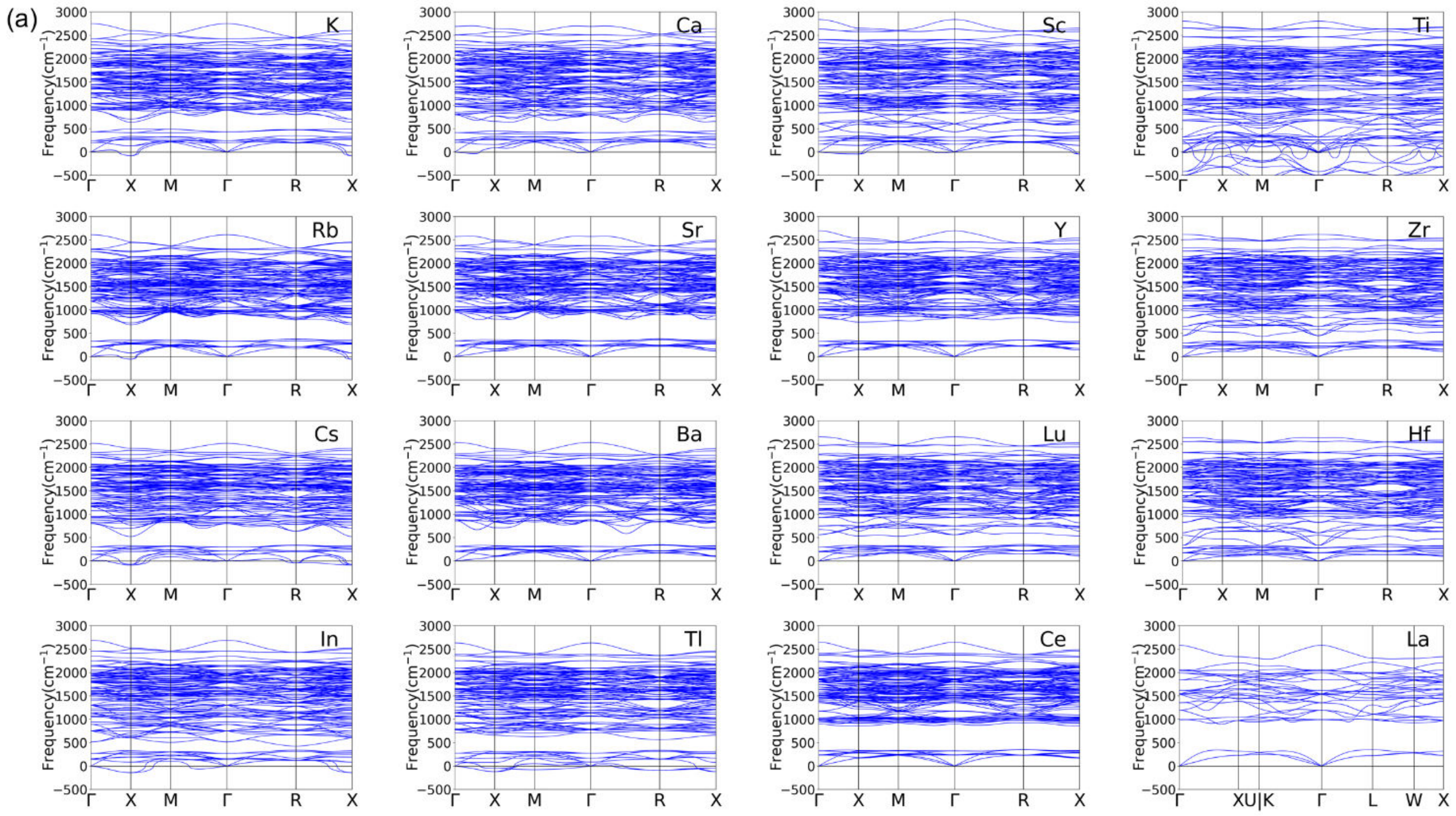

Fig. S5 (a) Phonon spectrum of $La_{0.75}M_{0.25}H_{10}$ at 400GPa. For example, 'K' means $La_{0.75}K_{0.25}H_{10}$.

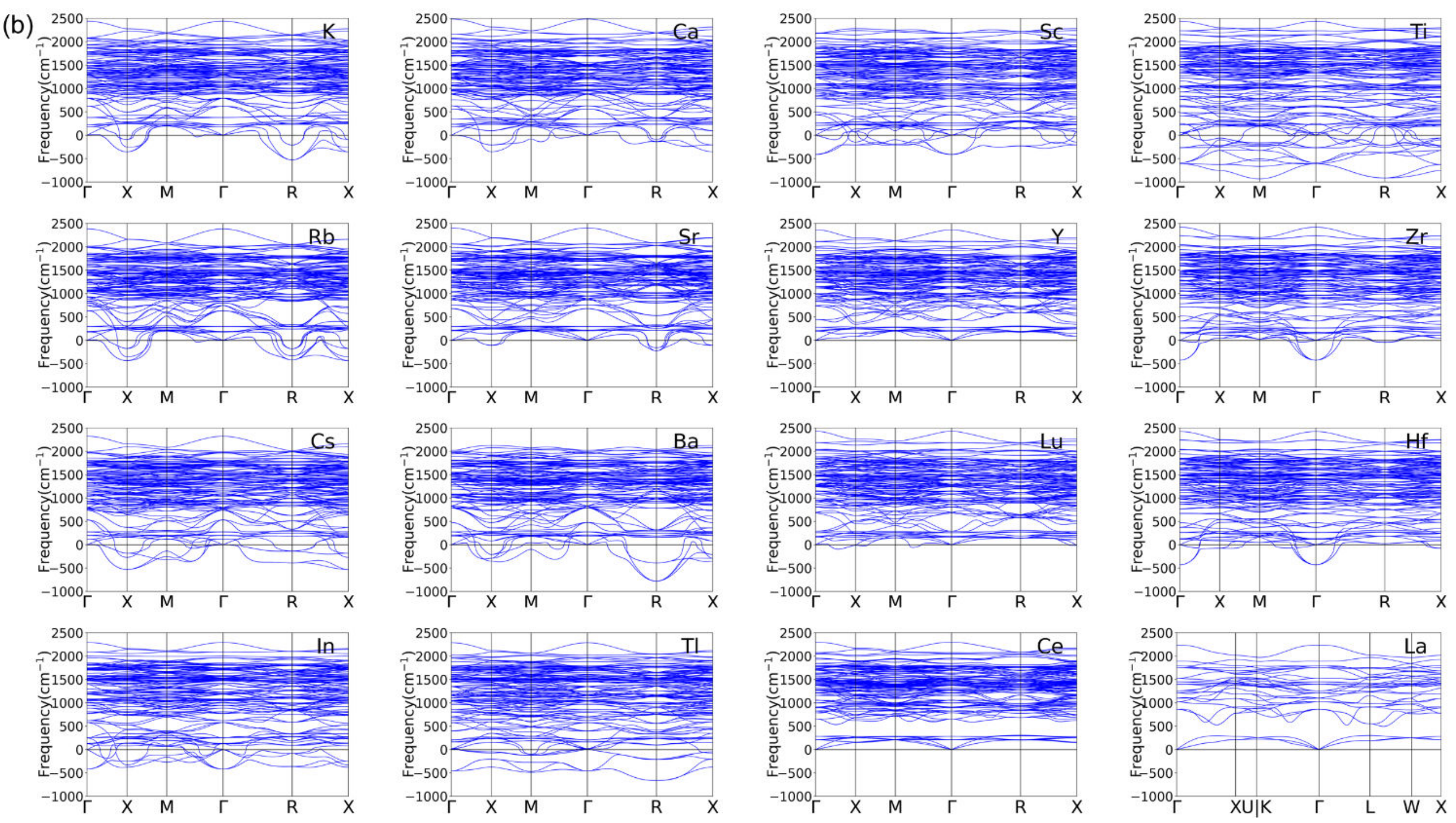

Fig. S5 (b) Phonon spectrum of $La_{0.75}M_{0.25}H_{10}$ at 250GPa

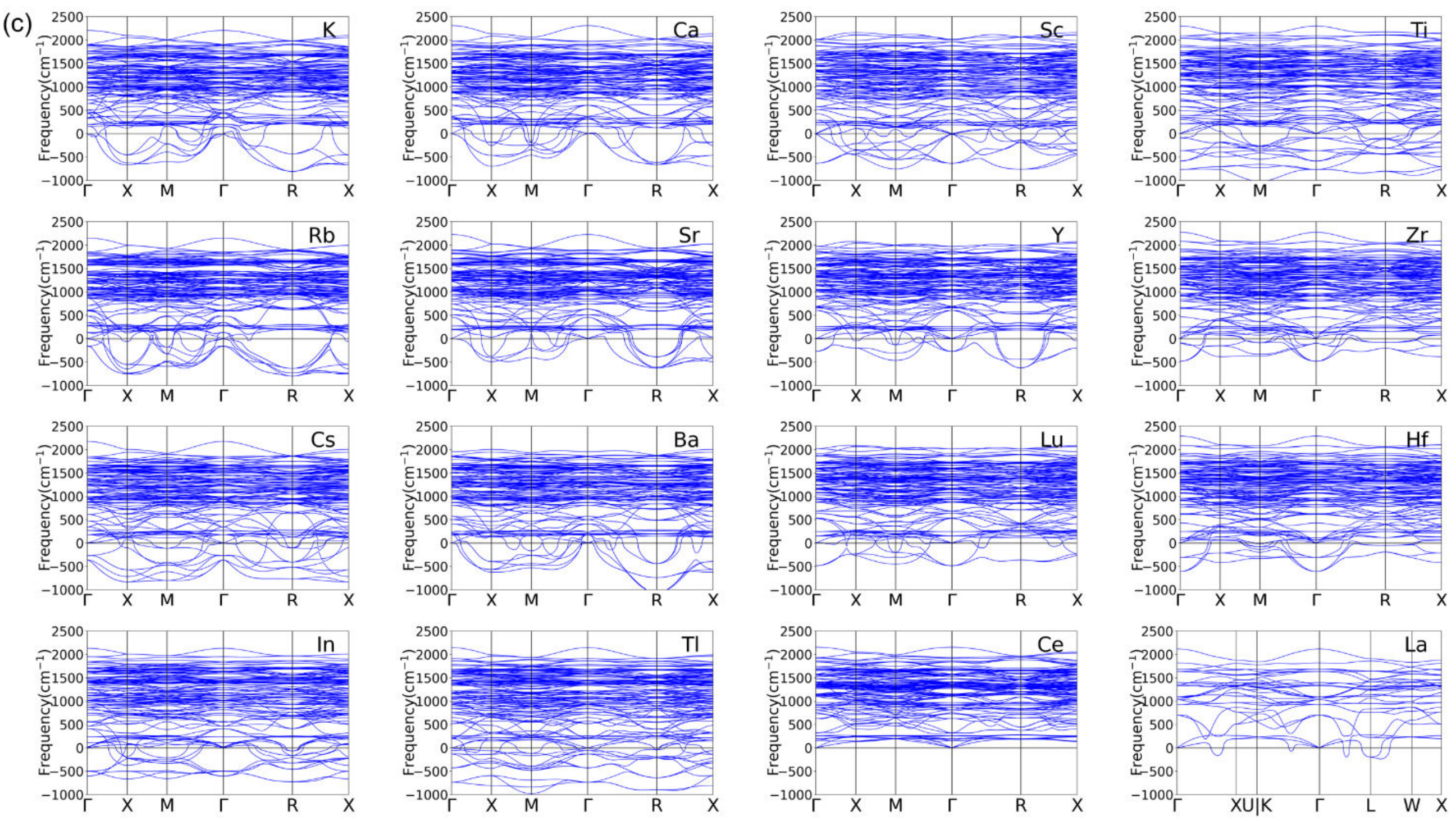

Fig. S5 (c) Phonon spectrum of $La_{0.75}M_{0.25}H_{10}$ at 200GPa.

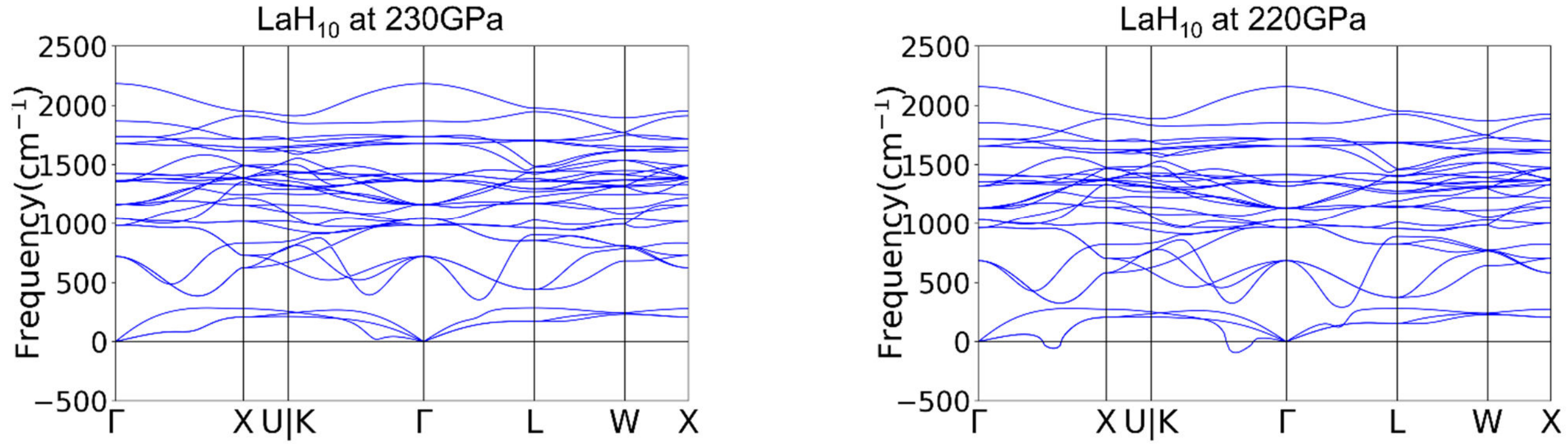

Fig. S6. Harmonic phonon spectrum of $LaH_{10}$ at 230GPa and 220GPa.

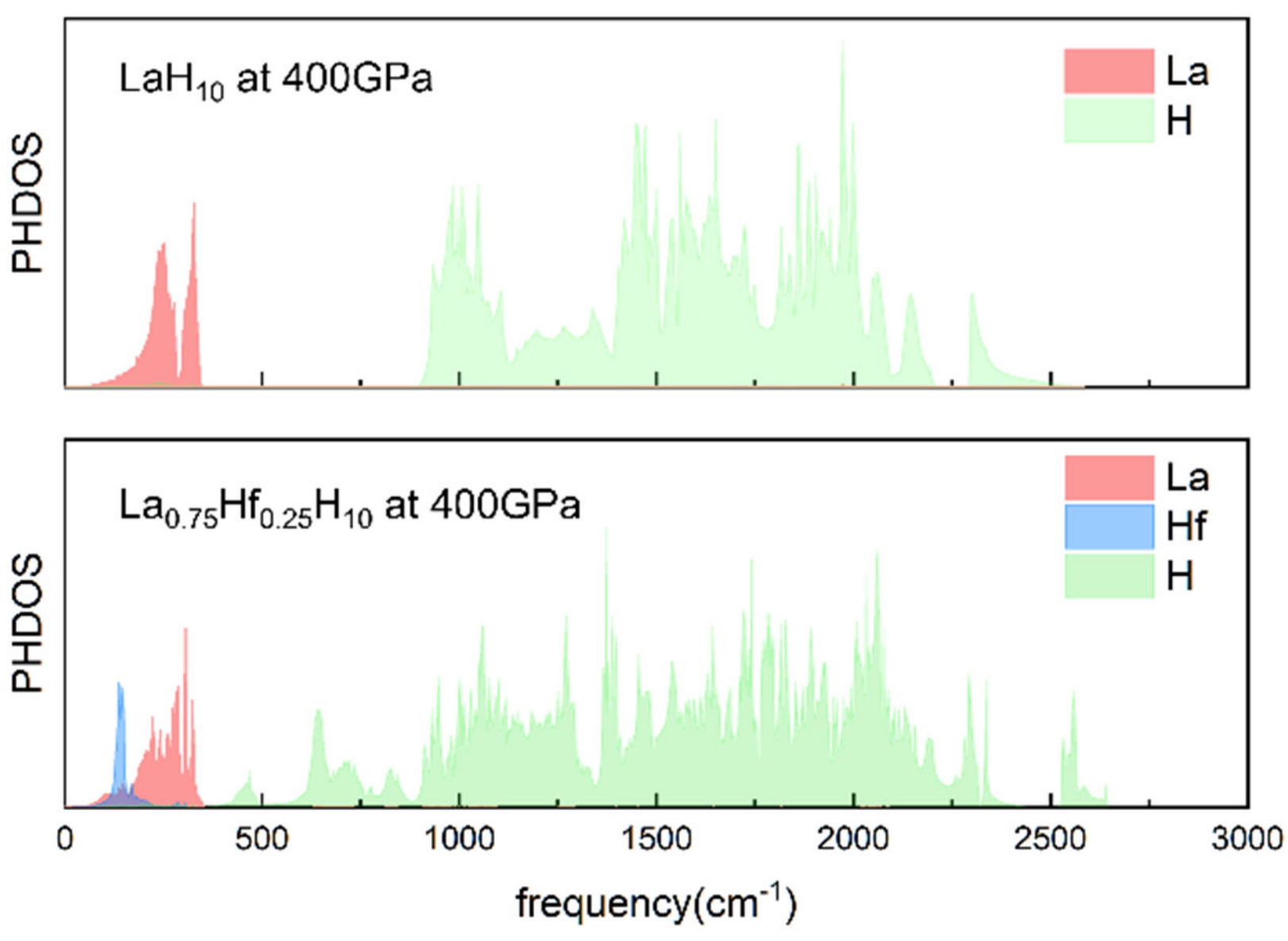

Fig. S7. Projected phonon density of states of $LaH_{10}$ and $La_{0.75}Hf_{0.25}H_{10}$ at 400GPa.

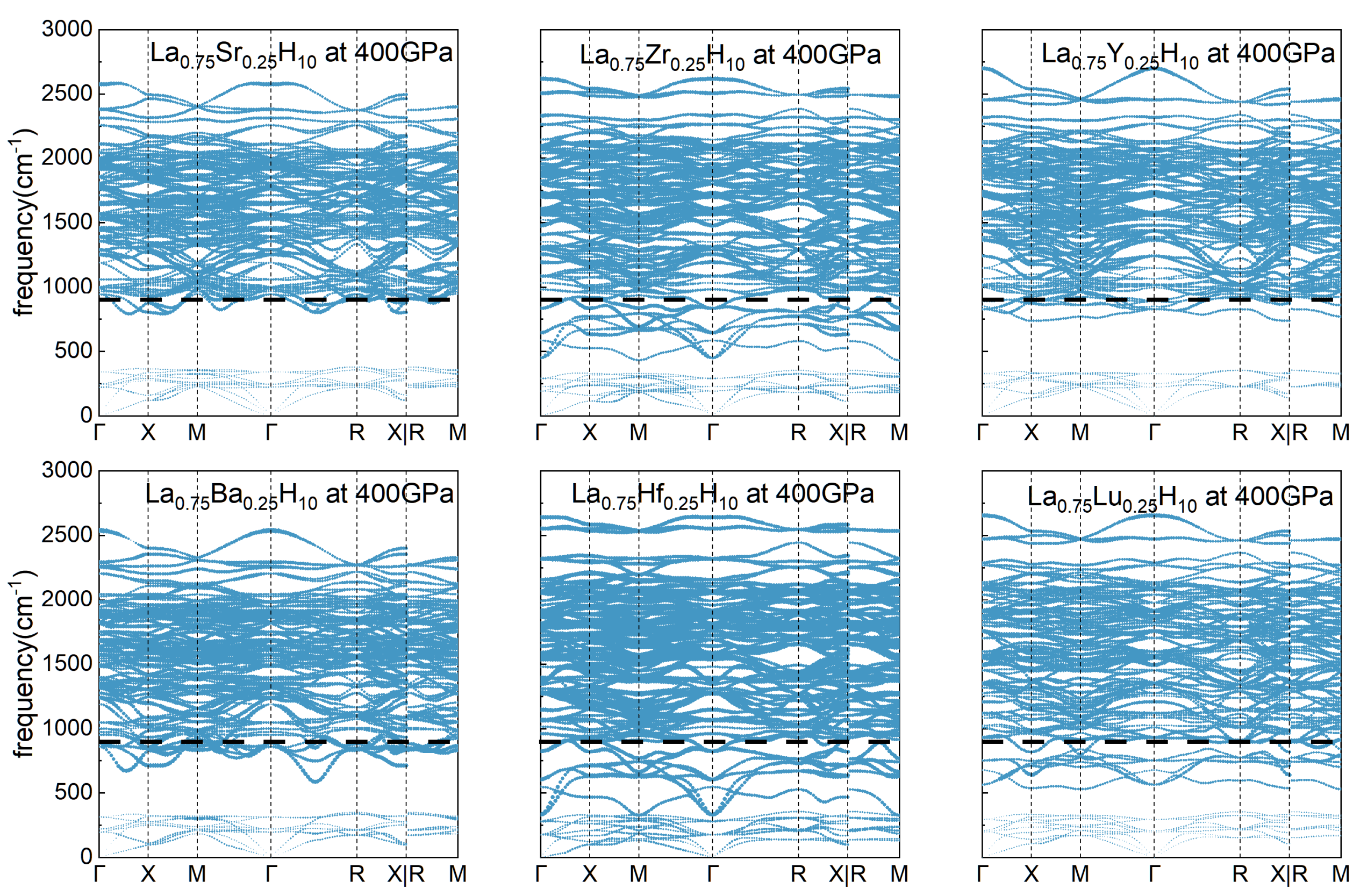

Fig. S8. Phonon spectrum of $La_{0.75}M_{0.25}H_{10}$ (M=Sr, Ba, Zr, Hf, Y and Lu) at 400GPa. The solid circles show the EPC with the area proportional to the respective phonon linewidth. Their EPC constant are showed in TABLE. S2.

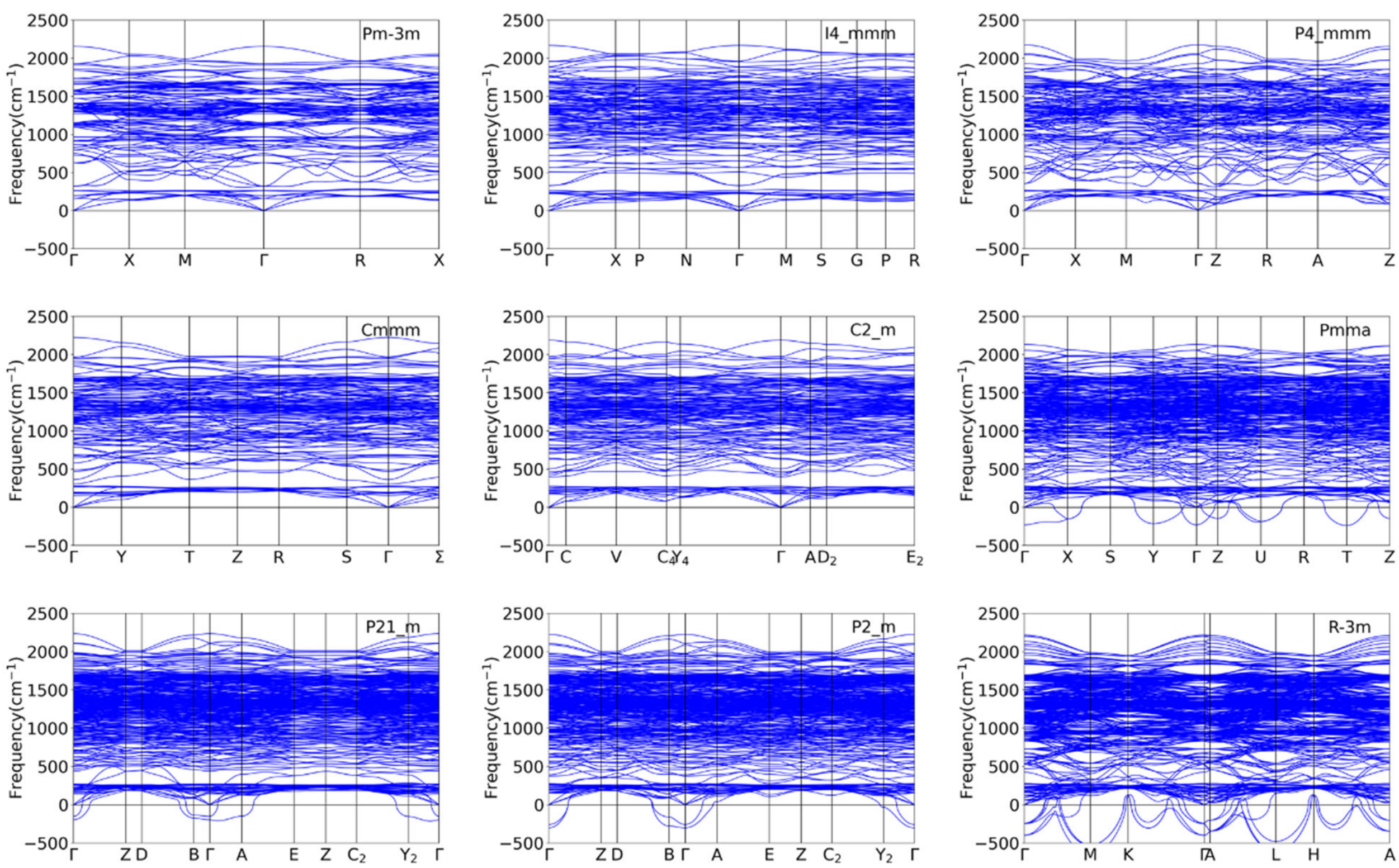

Fig. S9. Harmonic phonon spectrum of 9 $La_{0.75}Ce_{0.25}H_{10}$ polymorphs at 200GPa.

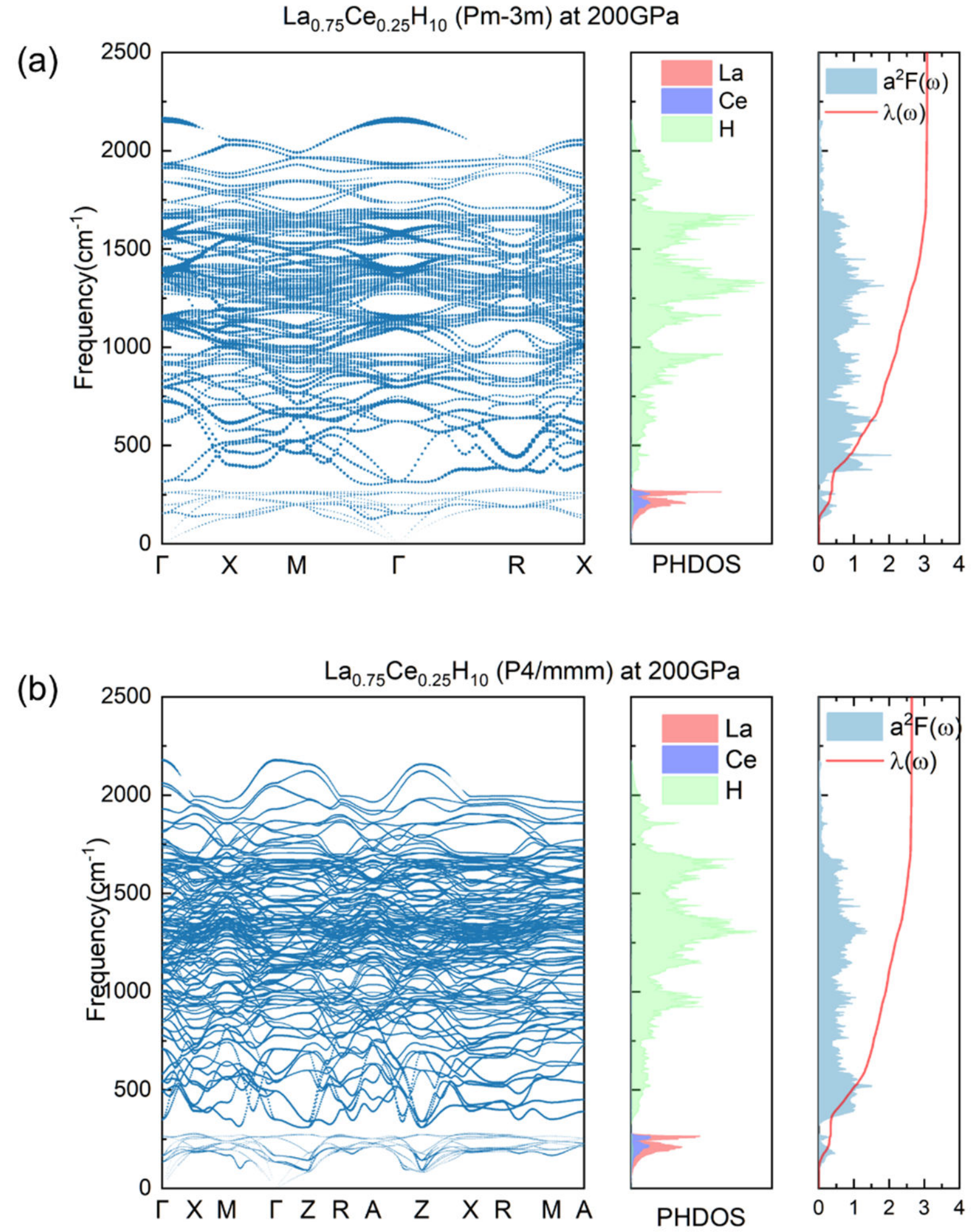

Fig. S10. Phonon spectrum, projected phonon DOS, Eliashberg spectrum function $\alpha^2F(\omega)$, and electron-phonon coupling integral $\lambda(\omega)$ of $La_{0.75}Ce_{0.25}H_{10}$ at 200GPa. (a) phase Pm-3m. (b) phase P4/mmm. The solid circles in the phonon spectrum show the EPC with the area proportional to the respective phonon linewidth.

Table S1. Valence electron configuration of 17 elements for PAW potential (ultrasoft pseudopotential) used in VASP (QE)

| **element** | **Valence configuration** | **element** | **Valence configuration** |
|---|---|---|---|
| H | $1s^1$ | Y | $4s^2 4p^6 4d^1 5s^2$ |
| La | $5s^2 5p^6 5d^1 6s^2$ | Ti | $3s^2 3p^6 3d^2 4s^2$ |
| K | $3s^2 3p^6 4s^1$ | Zr | $4s^2 4p^6 4d^2 5s^2$ |
| Rb | $4s^2 4p^6 5s^1$ | Hf | $5s^2 5p^6 5d^2 6s^2$ |
| Cs | $5s^2 5p^6 6s^1$ | Ce | $5s^2 5p^6 4f^1 5d^1 6s^2$ |
| Ca | $3s^2 3p^6 4s^2$ | Lu | $5s^2 5p^6 4f^{14} 5d^1 6s^2$ |
| Sr | $4s^2 4p^6 5s^2$ | In | $4d^{10} 5s^2 5p^1$ |
| Ba | $5s^2 5p^6 6s^2$ | Tl | $5d^{10} 6s^2 6p^1$ |
| Sc | $3s^2 3p^6 3d^1 4s^2$ | | |

TABLE S2. λ(900cm$^{-1}$), λ, $\omega_{log}$ (K) of 7 structures at 400GPa. $\lambda(900\text{cm}^{-1}) = 2\int_0^{900cm-1} \frac{\alpha^2 F(\omega)}{\omega} d\omega$, representing for the contribution of phonon modes with frequency below 900cm$^{-1}$.

| **structure** | **λ**(900cm$^{-1}$) | **λ** | $\boldsymbol{\omega_{log}}$ |
|---|---|---|---|
| $LaH_{10}$ | 0.18 | 1.41 | 1619 |
| $La_3SrH_{40}$ | 0.43 | 1.69 | 1380 |
| $La_3BaH_{40}$ | 0.98 | 2.34 | 915 |
| $La_3ZrH_{40}$ | 0.98 | 2.34 | 1112 |
| $La_3HfH_{40}$ | 1.01 | 2.32 | 1038 |
| $La_3YH_{40}$ | 0.28 | 1.55 | 1557 |
| $La_3LuH_{40}$ | 0.54 | 1.73 | 1302 |